\documentclass[twocolumn,letter]{jpsj3}
\usepackage{txfonts}
\usepackage{bm}

\title{Possible Evolution of Antiferromagnetism in Zn-Doped Heavy-Fermion Superconductor CeCoIn$_{\bm 5}$}

\author{Makoto~Yokoyama,$^1$\thanks{E-mail address: makotti@mx.ibaraki.ac.jp} 
Kenji~Fujimura,$^1$ Sara~Ishikawa,$^1$ Masashi~Kimura,$^1$ Takeshi~Hasegawa,$^1$ Ikuto~Kawasaki,$^2$, Kenichi~Tenya$^3$, Yohei~Kono,$^4$ and Toshiro~Sakakibara$^4$}
\inst{$^1$Faculty of Science, Ibaraki University, Mito 310-8512, Japan\\
$^2$RIKEN Nishina Center for Accelerator Based Science, RIKEN, Wako 351-0198, Japan\\
$^3$Faculty of Education, Shinshu University, Nagano 380-8544, Japan\\
$^4$Institute for Solid State Physics, University of Tokyo, Kashiwa 277-8581, Japan
} 

\abst{
We have succeeded in growing single crystals of the heavy-fermion superconductor CeCo(In$_{1-x}$Zn$_x$)$_5$ with $x\le 0.07$. Measurements of specific heat, electrical resistivity, dc magnetization and ac susceptibility revealed that the superconducting (SC) transition temperature $T_{\rm c}$ decreases from 2.25 K ($x=0$) to 1.8 K ($x=0.05$) by doping Zn into CeCoIn$_5$. Furthermore, these measurements indicate a development of a new ordered phase below $T_{\rm o}\sim 2.2\ {\rm K}$ for $x\ge 0.05$, characterized by the reduced magnetization and electrical resistivity in the ordered phase, and the enhancement of specific heat at $T_{\rm o}$. This phase transition can be also recognized by the shoulder-like anomaly seen at $H_{\rm o}\sim 55\ {\rm kOe}$ in the field variations of the magnetization at low temperatures, which is clearly distinguished from the superconducting critical fields $H_{\rm c2}=49\ {\rm kOe}$ for $x=0.05$ and 42 kOe for $x=0.07$. We suggest from these results that the antiferromagnetic (AFM) order is generated by doping Zn, and the interplay between the SC and AFM orders is realized in CeCo(In$_{1-x}$Zn$_x$)$_5$. 
}

\begin{document}
\maketitle
The nature of unusually enhanced fluctuations and ordered states related to the quantum critical phenomena has been attracting much interest in the physics of the heavy-fermion systems. These features are observed in the vicinity of the quantum critical point (QCP), corresponding to the phase transition at zero temperature, which is generated by suppressing the magnetically ordered state via applying pressure, magnetic field and doping ions. Near the QCP, the non-Fermi-liquid (NFL) properties evolve in temperature variations of thermodynamic and transport quantities, indicating the existence of new types of low-energy excitations different from those expected in usual Fermi-liquid state. Furthermore, the unconventional superconductivity (SC) often emerges near the QCP, and it is therefore believed that the magnetic fluctuation enhanced near the QCP plays a crucial role in the formation of the Cooper pairs.

Among the heavy-fermion systems showing the quantum critical phenomena, the heavy-fermion superconductor CeCoIn$_5$ (the HoCoGa$_5$-type tetragonal structure) is one of the most intensively investigated compounds. The SC transition of CeCoIn$_5$ is characterized by an anomalously large specific-heat jump $\Delta C/\gamma T_{\rm c}=4.5$ at the transition temperature $T_{\rm c}=2.3\ {\rm K}$\cite{rf:Petrovic2001}. A strong Pauli-limited effect gives rise to a first-order transition at the SC critical field $H_{\rm c2}$ below 0.7 K\cite{rf:Izawa2001,rf:Tayama2002,rf:Ikeda2001}. In addition, above $H_{\rm c2}$ the NFL behavior is observed in the temperature variations of the bulk quantities, which is considered to be due to the effect of quantum critical fluctuation induced in the vicinity of antiferromagnetism\cite{rf:Bianchi2003}.

In fact, an antiferromagnetic (AFM) order is generated by substituting ions for Co and In. In the mixed compounds CeRh$_{1-x}$Co$_x$In$_5$\cite{rf:Zapf2001}, it is revealed that substituting Rh for Co reduces $T_{\rm c}$ down to $\sim 1.5\ {\rm K}$ at $x\sim 0.8$, and the AFM order develops in the Co concentrations below $x_{\rm c}\sim 0.8$. A continuous increase of the AFM transition temperature $T_{\rm N}$ with further decreasing $x$ suggests that the AFM-QCP exists at $x_{\rm c}$. A similar relationship between the SC and AFM orders is realized in the CeCo(In,Cd)$_5$ system\cite{rf:Pham2006,rf:Urbano2007,rf:Tokiwa2010,rf:Capan2010}. Doping Cd into CeCoIn$_5$ is found to monotonically suppress the SC phase and then generate the AFM order above a few percent of Cd concentration, where $T_{\rm N}$ seems to continuously increase from zero temperature with increasing Cd concentration. The substitution of Hg for In yields very similar change from SC to AFM states\cite{rf:Bauer2008}. In these alloys, the AFM order induced near the QCP involves a $q=(1/2,1/2,1/2)$ structure.\cite{rf:Ohira-Kawamura2007,rf:Yoko2006,rf:Nicklas2007,rf:Yoko2006-2} Furthermore, the coupling between the SC and the AFM is inferred from the observations of the spin-resonance excitation at the same $q$ vector in the SC phase of pure CeCoIn$_5$\cite{rf:Stock2008,rf:Raymond2012}. An exceptional substitution effect against these features is seen in the CeCo(In,Sn)$_5$ alloys\cite{rf:Bauer2006}, where doping Sn stabilizes a paramagnetic NFL state as a consequence of a suppression of the SC, and the AFM order has never observed throughout the variation of the ground states. 

In the present study, we have succeeded in growing new mixed compounds CeCo(In,Zn)$_5$ for the first time. In order to derive the characteristics of the SC phase, and find the relationship between SC and AFM instabilities, we have investigated low-temperature ordered states of CeCo(In,Zn)$_5$ by performing thermal, magnetic and transport experiments.   

Plate-shaped single crystals of CeCo(In$_{1-x}$Zn$_x$)$_5$ for $x=0,\ 0.025,\ 0.05$ and $0.07$, whose basal planes are perpendicular to the tetragonal $c$ axis, were grown by means of the In flux technique. High-purity materials of Ce, Co, Zn, and In were set and sealed under 0.02 MPa Ar atmosphere in a quartz tube, where we selected a molar ratio of the materials to be ${\rm Ce:Co:In:Zn}=1:1:F\times (1-x):F\times x$ with $F=20-25$. We find it very hard to grow the crystal as $x$ is increased, perhaps caused by an absence of the end material CeCoZn$_5$ as a stable composition\cite{rf:Makaryk2001}. We checked by means of the x-ray diffraction that these samples have tetragonal crystal structure. The energy dispersive x-ray spectrometry (EDS) measurements for the samples used in the present experiments indicate that the estimated Zn concentrations $y$ significantly deviate from the nominal values $x$, though the elements homogeneously distribute in the sample. It is known that such a deviation frequently occurs in the sample grown by the flux technique. The $y$ values are estimated to be roughly 20\% of $x$, which seem to be linked with the reduction of nearly the same amount of In, and uncorrelated with the concentrations of the other elements. However, it is hard to determine exact $y$ values using the EDS technique for the smallness of $y$. This situation is quite similar to the report on the CeCo(In,Cd)$_5$ system\cite{rf:Pham2006}. At present, we consider from these analyses that our crystals have the CeCo(In,Zn)$_5$ formula. This may be supported by the facts that both Zn and Cd belong to the same group, and it is concluded from the microscopic investigations that the Cd ions occupy the crystallographic sites equivalent to those of In in CeCo(In,Cd)$_5$\cite{rf:Pham2006}. In addition, we use the nominal $x$ value throughout this paper for clarity and simplicity.

The electrical resistivity measurements were performed with a standard four-wire technique from 1.0 K to room temperature. The dc-magnetization measurements were carried out using a commercial SQUID magnetometer (MPMS) in the temperature range of $2-300\ {\rm K}$ and magnetic fields $H$ up to 50 kOe. We also measured dc magnetization using a high-precision capacitive magnetometer based on the Faraday-force technique\cite{rf:Sakakibara94} between 0.27 K and 3 K in fields up to 80 kOe. The $a$ axis ac-susceptibility measurements were performed between 1.2 K and 4 K using a standard Hartshorn bridge method, where ac field is applied parallel to the basal plane of the plate-shaped samples to reduce a demagnetizing-field effect. Magnitude and frequency of applied ac-field were selected to be $\sim0.5\ {\rm Oe}$ and 180 Hz. Specific heat was measured using a thermal relaxation method from 1.1 K to 10 K.  

\begin{figure}[tbp]
\begin{center}
\includegraphics[bb=82 436 426 813,keepaspectratio,width=0.4\textwidth]{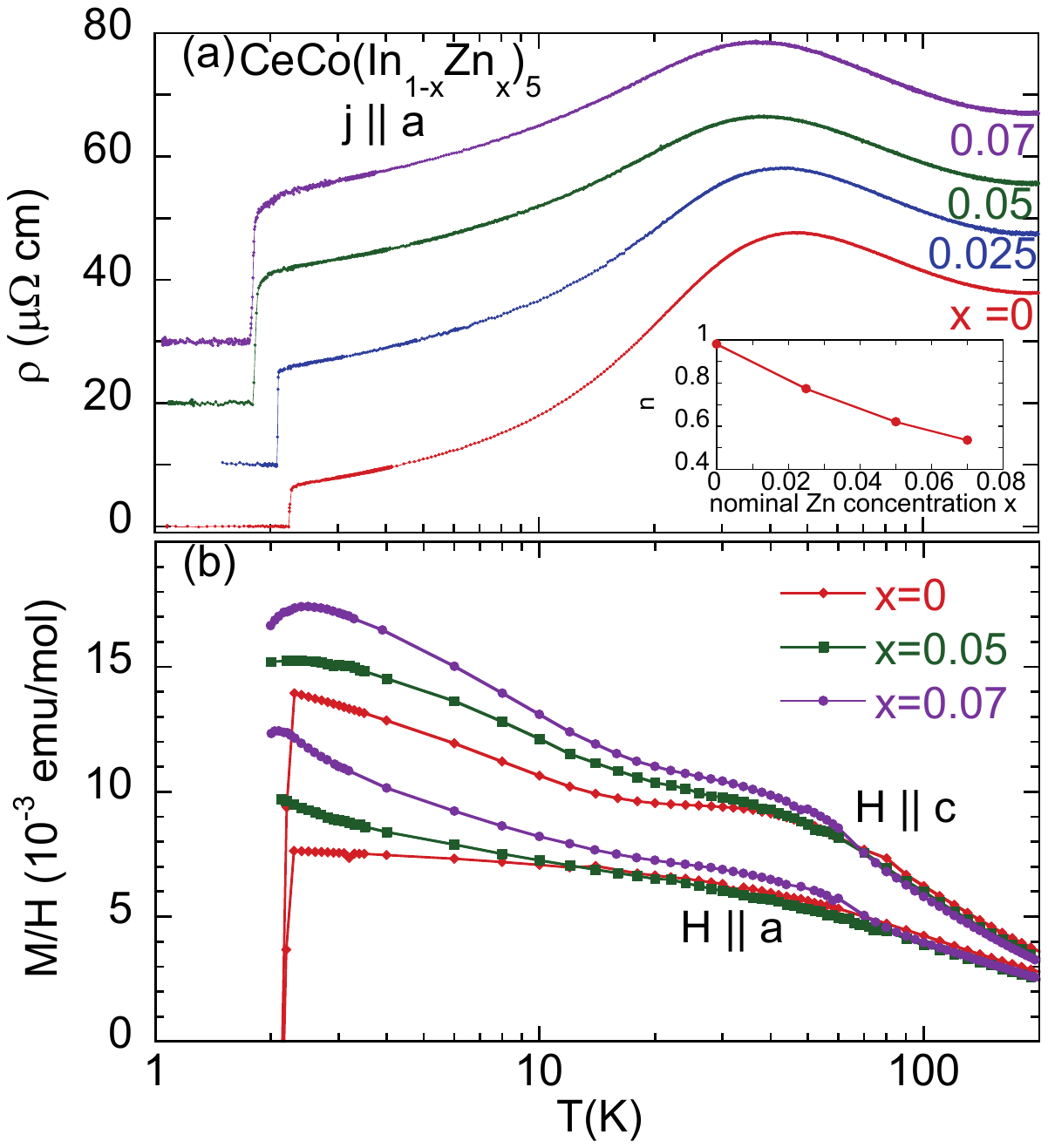}
\end{center}
  \caption{
(Color online) (a) Temperature variations of $a$-axis electrical resistivity $\rho(T)$ for CeCo(In$_{1-x}$Zn$_x$)$_5$ with $x \le 0.07$. Note that the baselines of the data for $x \ge 0.025$ are transformed for clarity. The inset shows the $n$ parameter in the $T^n$ function of $\rho(T)$, obtained by fitting the $\rho(T)$ data between 3 K and 20 K. (b) Temperature variations of magnetic susceptibility derived from dc magnetization divided by magnetic field of $H \le 5\ {\rm kOe}$. 
}
\end{figure}
Figure 1(a) shows temperature variations of $a$-axis electrical resistivity $\rho(T)$ for CeCo(In$_{1-x}$Zn$_x$)$_5$. The zero resistivity due to the SC transition is observed in $\rho(T)$ for all the $x$ range presently investigated. For $x=0$, as is reported previously,\cite{rf:Petrovic2001} $\rho(T)$ linearly increases with increasing temperature above $T_{\rm c}=2.25\ {\rm K}$, followed by an appearance of a broad maximum at $T_{\rm max}\sim 47\ {\rm K}$. In the Zn-doped samples, the $\rho(T)$ curve for $T_{\rm c} < T <T_{\rm max}$ changes into a convex behavior in the linear temperature scale, and $T_{\rm max}$ decreases to $\sim 37\ {\rm K}$ at $x=0.07$. In order to figure out the former variation, we assume that $\rho(T)$ is proportional to $T^n$ for $T_{\rm c} < T <T_{\rm max}$, and shown in the inset of Fig.\ 1(a) is the $n$ parameter estimated from the best fit to the $\rho(T)$ data between 3 K and 20 K. A large deviation of the $\rho(T)$ curve from the $T^2$ function indicates that electron transports involve the non-Fermi liquid characteristics in the present systems. It is further found that the $n$ parameter are reduced from $\sim 1$ ($x=0$) to $\sim 0.5$ ($x=0.7$) with increasing $x$, suggesting that doping Zn changes the mechanism of excitations and scatterings of the conduction electrons above $T_{\rm c}$.

In Fig.1(b), we plot temperature variations of $c$ and $a$ axis magnetic susceptibilities $M/H$, obtained by the SQUID magnetometer. For both directions, $M/H$ exhibits a shoulder-like behavior at nearly the same temperature as $T_{\rm max}$ defined in $\rho(T)$. At high temperatures above $T_{\rm max}$, $M/H$ obeys fairly well the Curie-Weiss law $N\mu_{\rm eff}^2/3k_{\rm B}(T-\theta_{\rm p})$. The best fits in the temperature range of $100-300\ {\rm K}$ for $x=0$ provide the $\mu_{\rm eff}$ and $\theta_{\rm p}$ values of 2.57(5) $\mu_{\rm B}/{\rm Ce}$ and $-97(4)$ K for $a$ axis, and 2.58(2) $\mu_{\rm B}/{\rm Ce}$ and $-31(3)$ K for $c$ axis, respectively, and the similar values are obtained for the Zn-doped samples. The $\mu_{\rm eff}$ values are comparable to 2.54 $\mu_{\rm B}$ expected from the $J=5/2$ multiplet in the Ce$^{3+}$ ion, and the negative $\theta_{\rm p}$ reflects an existence of an AFM correlation in these compounds. At low temperatures, we observed a sudden drop of $M/H$ for $x=0$, attributed to a shielding effect of the SC order. The diamagnetic signal is invisible in the $M/H$ data above $x=0.05$ because $T_{\rm c}$ becomes lower than the lowest accessible temperature of the SQUID magnetometer (2.0 K). Instead, a cusp is found in the magnetic susceptibility at $\sim 2.3\ {\rm K}$ for $x=0.07$.

\begin{figure}[tbp]
\begin{center}
\includegraphics[bb=65 75 456 780,keepaspectratio,width=0.4\textwidth]{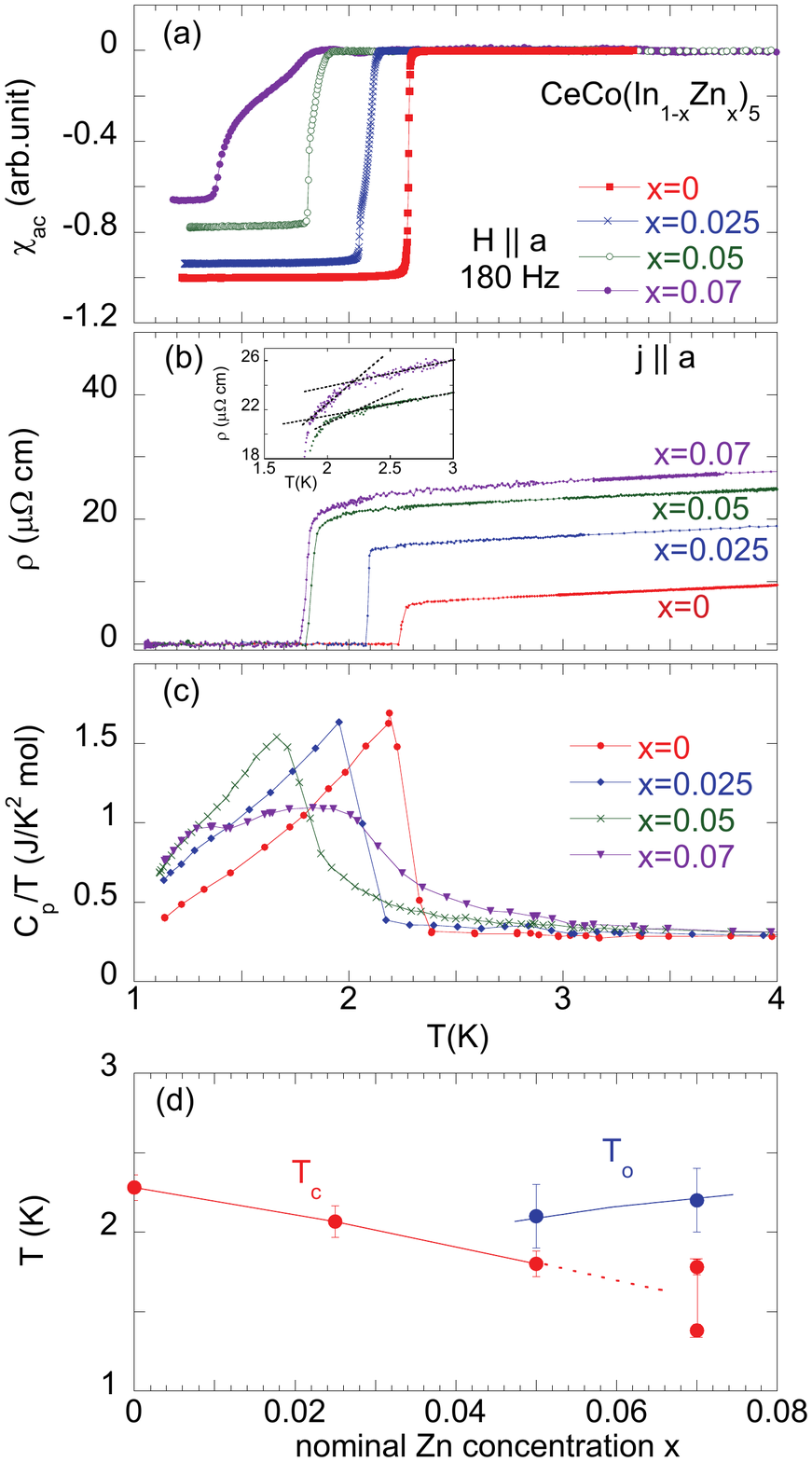}
\end{center}
  \caption{
(Color online) Temperature variations of (a) ac susceptibility, (b) electrical resistivity, and (c) specific heat divided by temperature for CeCo(In$_{1-x}$Zn$_x$)$_5$ ($x \le 0.07$), plotted in the vicinity of the SC transition temperatures. The inset of (b) shows an enlargement of the electrical resistivity for $x=0.05$ and $0.07$ around 2 K. The broken lines in the inset of (b) are a guide to the eye. (d) The $x-T$ phase diagram of CeCo(In$_{1-x}$Zn$_x$)$_5$ ($x \le 0.07$) obtained from temperature variations of ac susceptibility, electrical resistivity, and specific heat. 
}
\end{figure}
Displayed in Figs.\ 2(a)-(c) are the properties of ac susceptibility $\chi_{\rm ac}$, electrical resistivity $\rho$, and specific heat divided by temperature $C_p/T$ at $\sim T_{\rm c}$. For pure CeCoIn$_5$, the SC transition at $T_{\rm c}=2.25\ {\rm K}$ is clearly indicated by a drop of resistivity toward zero, a sharp and large diamagnetic signal in $\chi_{\rm ac}$, and a discontinuous jump in $C_p/T$. Such features are also seen in the data for $x \le 0.05$, but $T_{\rm c}$ is reduced to $1.8\ {\rm K}$ at $x=0.05$. On the other hand, a distribution of the SC transition temperature in the sample is inferred for $x=0.07$ since diamagnetic signal gradually develops in $\chi_{\rm ac}$. The drop of resistivity occurs at $T_{\rm c}^+= 1.78\ {\rm K}$, which corresponds to the onset of the diamagnetism in $\chi_{\rm ac}$. Subsequently, a small peak observed in $C_p/T$ at $T_{\rm c}^-=1.36\ {\rm K}$ is related to a sharp drop of $\chi_{\rm ac}$. The specific-heat jump $\Delta C_p/\gamma T_{\rm c}$ associated with the SC transition decreases from 4.6 ($x=0$) to 3.8 ($x=0.025$), and then becomes $\sim 1.7$ ($x=0.05$), implying that the SC order weakens and then occurs inhomogeneously in the Zn-doped samples. It may be supported by the reduced diamagnetic magnitude of $\chi_{\rm ac}$ with doping Zn, though the absolute values of $\chi_{\rm ac}$ involve fairly large errors (roughly $20\%$) in the present experiment because the Zn-doped samples are small and their shapes cannot be exactly determined.

Interestingly, $C_p/T$ exhibits a characteristic upturn below $\sim 2.1\ {\rm K}$ ($=T_{\rm o}$) for $x=0.05$. Moreover, a large shoulder-like anomaly develops in $C_p/T$ at $T_{\rm o}\sim 2.2\ {\rm K}$ for $x=0.07$, apart from the small peak at $T_{\rm c}^-=1.36\ {\rm K}$. We here define $T_{\rm o}$ as a midpoint of the $C_p/T$ increase. At $\sim T_{\rm o}$, the cusp appears in temperature variations of $M/H$ along $c$ and $a$ axes for $x=0.07$, and the saturation in the $c$ axis $M/H$ for $x=0.05$. In addition, $\rho(T)$ for $x=0.05$ and $0.07$ shows a weak kink at $\sim T_{\rm o}$ (the inset of Fig.\ 2(b)). These anomalies occur at higher temperatures than $T_{\rm c}$ (or $T_{\rm c}^+$ and $T_{\rm c}^-$ for $x=0.07$) determined by $\chi_{\rm ac}$ and $\rho(T)$, and hence, it is likely that they are the signatures of a phase transition or fluctuation other than the SC transition. It should be stressed that the anomaly at $T_{\rm o}$ will not be ascribed to the fragmentary SC order caused by the inhomogeneity of the samples. If the SC ordering occurs at $T_{\rm o}$ in parts of the sample, $\rho(T)$ is generally expected to show a large and discontinuous drop at the corresponding temperature, since a pass for the SC current may be generated by the fragmentary SC regions being linked with each other in the sample. $\chi_{\rm ac}$ is also expected to markedly go down below $T_{\rm o}$. However, present $\rho(T)$ and $\chi_{\rm ac}$ do not exhibit such behavior, whereas the relatively large anomaly evolves at $\sim T_{\rm o}$ in $C_p/T$ as an extensive variable. The entropy changes between 1.1 K and 3.0 K (above $T_{\rm c}$ and $T_{\rm o}$) are estimated to be 0.24(1)$R\ln 2$ for all the Zn concentrations presently investigated, but we cannot estimate the entire entropy change associated with the low-temperature orders for a lack of the $C_p/T$ data below 1 K. In Fig.\ 2(d), we plot the $x-T$ phase diagram obtained from the above estimations.

The low-temperature $c$-axis magnetization is also measured using the capacitive Faraday-force magnetometer. For the Zn-doped samples, a very large hysteresis loop associated with the SC order appears in the magnetization curves measured under increasing and decreasing field variations. It is considered that the large hysteresis in the magnetization curves reflects an effect of disorder induced by doping. Figure 3(a) shows magnetization divided by field $M/H$ at 0.27 K and around $H_{\rm c2}$ for the Zn-doped samples, plotted as a function of $H$. $H_{\rm c2}$ can be recognized by the close of hysteresis loop in $M/H$, corresponding to the fields of 50.5 kOe, 49 kOe and 42 kOe for $x=0.025$, $0.05$ and $0.07$, respectively. The details of the $M/H$ properties at $H_{\rm c2}$ for $x=0.025$ cannot be determined due to an occurrence of a peak effect just below $H_{\rm c2}$. On the other hand, the continuous variations of $M/H$ at $H_{\rm c2}$ for $x\ge 0.05$ suggest that the SC order breaks as a second order phase transition, in contrast to the first-order transition indicated in pure CeCoIn$_5$\cite{rf:Izawa2001,rf:Tayama2002,rf:Ikeda2001}. Above $H_{\rm c2}$, $M/H$ increases with increasing $H$, and then shows kink or shoulder-like anomalies at $H_{\rm o}\sim 55\ {\rm kOe}$ for both $x=0.05$ and $0.07$, followed by the constant magnitude of $M/H$ above $H_{\rm o}$. This feature for $H \le H_{\rm o}$ is clearly different from that observed in pure\cite{rf:Tayama2002,rf:Bianchi2003,rf:Ronning2005} and Zn 2.5\% doped CeCoIn$_5$ as well as usual paramagnetic responses, and hence suggests an existence of an ordered phase involving the suppressed uniform magnetization below $H_{\rm o}$. It is also found that such a reduced magnetization below $H_{\rm o}$ accompanies a peak structure in temperature variations of $M/H$ for $H < H_{\rm o}$ (Fig.\ 3(b)). For $H > H_{\rm o}$, by contrast, $M/H$ exhibits a monotonic and weak growth with decreasing temperature, quite similar to the observations above $H_{\rm c2}$ in pure CeCoIn$_5$\cite{rf:Tayama2002,rf:Bianchi2003,rf:Ronning2005}. In Fig.\ 3(c) we show the $H-T$ phase diagram for $x=0.05$, obtained from the temperature and field variations of $M/H$. For comparison, $H_{\rm c2}$ of pure CeCoIn$_5$\cite{rf:Tayama2002} is also plotted. One remarkable feature in this diagram is that the magnitude of $H_{\rm c2}$ at low temperatures for $x=0.05$ coincides with that of CeCoIn$_5$, though $T_{\rm c}$ for $x=0.05$ is significantly reduced. We find that $H_{\rm c2}$ is robust for the substitutions of Zn ions less than $x=0.05$ (Fig.\ 3(d)), while $T_{\rm c}$ monotonically decreases with increasing $x$ (Fig.\ 2(d)). An another one is that the $H_{\rm o}$ boundary encloses the SC phase region in the $H-T$ phase diagram, but its shape is not simply analogous with the SC boundaries for both pure and Zn 5\% doped CeCoIn$_5$. This suggests that the ordered phase below $H_{\rm o}$ involves different characters of stability against temperature and field from those of the SC order. Remarkably, at $H\sim 0$ the phase boundary concerning $H_{\rm o}$ agrees with $T_{\rm o}$ determined by the bulk quantities at zero field (see Fig. 2), reflecting the same origin underlying in the phase transition at $T_{\rm o}$ and $H_{\rm o}$. We have verified that the characteristics of the magnetization and the $H-T$ phase diagram for $x=0.07$ are quite similar to those for $x=0.05$.     
\begin{figure}[tbp]
\begin{center}
\includegraphics[bb=19 343 491 728,keepaspectratio,width=0.43\textwidth]{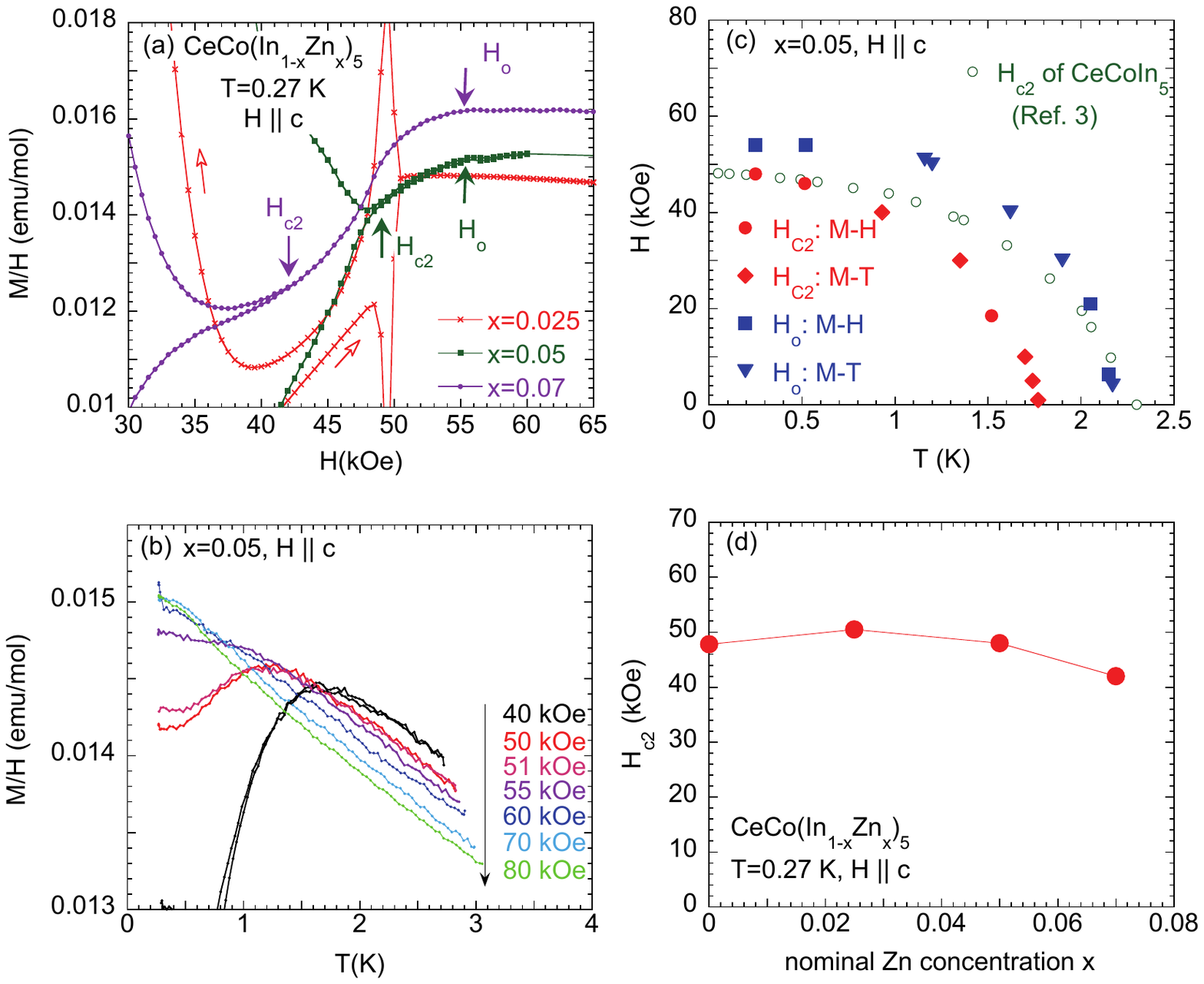}
\end{center}
  \caption{
(Color online) (a) The $c$ axis $M/H$ curve at 0.27 K and around $H_{\rm c2}$ for CeCo(In$_{1-x}$Zn$_x$)$_5$, plotted as functions of magnetic field. (b) $M/H$ versus temperature obtained under several magnetic fields comparable to $H_{\rm c2}$ for $x=0.05$. (c) The $H-T$ phase diagram for $x=0.05$. The $c$ axis $H_{\rm c2}$ for pure CeCoIn$_5$ is also plotted as a comparison\cite{rf:Tayama2002}. (d) The $c$ axis $H_{\rm c2}$ values at $0.27\ {\rm K}$ for $x \le 0.07$, plotted as a function of nominal Zn concentration $x$.
}
\end{figure}

It is found that the ordered phase other than the SC emerges below $T_{\rm o}$ in CeCo(In$_{1-x}$Zn$_x$)$_5$ for $x\ge 0.05$. This order can be rather realized by the kink or shoulder-like anomalies in the field variations of $M/H$, while very sharp and discontinuous jump associated with the ordering cannot be observed in the specific heat at zero field. We simply consider that the ordered state below $T_{\rm o}$ and $H_{\rm o}$ is ascribed to the AFM, since the magnitude of $M/H$ is suppressed in the ordered phase as compared with that in the paramagnetic phase. The evolution of the AFM order as well as the monotonic reductions of both $T_{\rm c}$ and $\Delta C_p/\gamma T_{\rm c}$ with increasing $x$ suggests that doping Zn makes a distance from the AFM-QCP located at $x\sim 0$. Though this situation, $\rho(T)$ at low temperatures does not show a tendency of the Fermi-liquid behavior. This may be related to an effect of fluctuating AFM moment above $T_{\rm o}$. Similar argument has been proposed for the CeRh$_{1-x}$Co$_x$In$_5$ system.\cite{rf:Jeffries2005} The appearance of the AFM order has also been found in the other substitutions for CeCoIn$_5$\cite{rf:Zapf2001,rf:Pham2006,rf:Bauer2008}. In particular, the AFM order is generated in the substitutions involving Cd and Hg that belong to the same group as Zn, and these $x-T$ phase diagrams are very similar to the presently revealed one for CeCo(In$_{1-x}$Zn$_x$)$_5$. At the same time, we presume that this AFM order may occur inhomogeneously in the sample, because the suppression of $M/H$ below $H_{\rm o}$ and the signatures of the transition at $T_{\rm o}$ and $H_{\rm o}$ are neither very sharp nor very large at least for $x=0.05$, although the static nature of the AFM order is indicated by the appearance of the bulk anomalies. It is probable that these features are related to the disorder induced by doping Zn. It is still unclear at present whether the SC and AFM orders are microscopically coexistent or spatially divided. For instance, it is pointed out by the NMR experiments that the AFM droplets are formed around the doped Cd ions in the CeCo(In,Cd)$_5$ system\cite{rf:Urbano2007}. 
 
As for the SC properties, the present investigation revealed that $H_{\rm c2}$ is robust for doping Zn in spite of the significant reduction of $T_{\rm c}$. This will be attributed to the strong Pauli-paramagnetic effect on the SC order involved in the present system, as is argued in pure CeCoIn$_5$\cite{rf:Izawa2001,rf:Tayama2002,rf:Ikeda2001}. For $x=0.05$, we can estimate from the slope of $H_{\rm c2}$ at $T_{\rm c}$ ($dH_{\rm c2}/dT=-128\ {\rm kOe/K}$) that the orbital-limited field of the SC pair breaking $H_{\rm c2}^{\rm orb}$ has a magnitude of 156 kOe for a dirty-limit assumption. The $H_{\rm c2}^{\rm orb}$ value is comparable to that suggested for pure CeCoIn$_5$ ($\sim 150\ {\rm kOe}$)\cite{rf:Tayama2002,rf:Ikeda2001}, and much larger than the Pauli-limited field $H_{\rm P}$ when the relation $H_{\rm P}\sim H_{\rm c2}$ is assumed. We thus consider that the spin-pair breaking mechanism governs the violation of the SC order at $H_{\rm c2}$ even in the Zn-doped samples. In this context, we expect that the competition between the reduced spin susceptibility due to the AFM ordering and the suppression of the SC condensation energy should affect the magnitude of $H_{\rm P}$, leading to the different $x$ dependence between $T_{\rm c}$ and $H_{\rm c2}$. To confirm the occurrence of the AFM ordering and then clarify the details on the interplay between the SC and AFM orders, we plan to perform further precise investigations using microscopic probes. 

In summary, our investigations on the low-temperature properties of CeCo(In$_{1-x}$Zn$_x$)$_5$ revealed that the SC transition temperature is reduced from 2.25 K ($x=0$) to 1.8 K ($x=0.05$) with increasing $x$, while $H_{\rm c2}$ along the $c$ axis is nearly independent of $x$ for $x\le 0.05$. On the other hand, the evolution of the ordered phase for $x\ge 0.05$ is suggested from the anomalies at $T_{\rm o}\sim 2.2\ {\rm K}$ in the bulk quantities, and also indicated by the kink or shoulder-like anomalies in the field variations of $M/H$. This can be ascribed to the AFM order because $M/H$ is reduced in this ordered phase. These results suggest that the interplay between the SC and AFM orders is generated in CeCo(In$_{1-x}$Zn$_x$)$_5$ for $x\ge 0.05$.   

\begin{acknowledgment}
We are grateful to S. Nakano and K. Kuwahara for technical support on the sample preparations. This work was partly supported by Grants-in-Aid for Scientific Research on Innovative Areas ``Heavy Electrons" and ``Topological Quantum Phenomena" from the Ministry of Education, Culture, Sports, Science and Technology of Japan. 
\end{acknowledgment}

\end{document}